# Malware Detection Approach for Android systems Using System Call Logs

*Sanya Chaba*
Computer Engineering
National Institute of Technology
Kurukshetra
Haryana, India
sanyachaba26@gmail.com

*Rahul Kumar*
Computer Engineering
National Institute of Technology
Kurukshetra
Haryana, India
rahulk0311@gmail.com

*Rohan Pant*
Computer Engineering
National Institute of Technology
Kurukshetra
Haryana, India
rohanpant26@gmail.com

*Mayank Dave*
Computer Engineering
National Institute of Technology
Kurukshetra
Haryana, India
mdave@nitkkr.ac.in

*Abstract---* Static detection technologies based on signature-based approaches that are widely used in Android platform to detect malicious applications. It can accurately detect malware by extracting signatures from test data and then comparing the test data with the signature samples of virus and benign samples. However, this method is generally unable to detect unknown malware applications. This is because, sometimes, the machine code can be converted into assembly code, which can be easily read and understood by humans. Furthuremore, the attacker can then make sense of the assembly instructions and understand the functioning of the program from the same. Therefore we focus on observing the behaviour of the malicious software while it is actually running on a host system. The dynamic behaviours of an application are conducted by the system call sequences at the end. Hence, we observe the system call log of each application, use the same for the construction of our dataset, and finally use this dataset to classify an unknown application as malicious or benign.

## I. INTRODUCTION

Malware or Malicious Software is defined as software designed to distort and interrupt the mobile or computer applications, collect important information and hence perform malicious operations. These malicious operations include gaining access over private information, covertly steal this valuable information over the system, display undesirable advertisement, and spy on the activities of the users. The following figure (Figure 1) represents different types of malware:

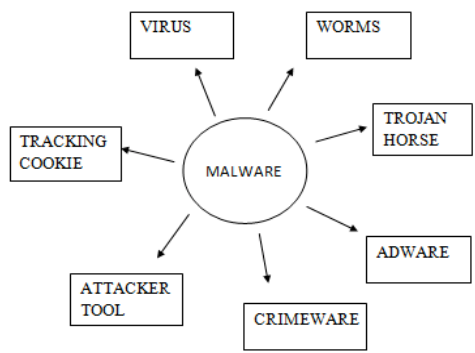

**Figure 1: Types of Malware**

Computer security has always posed serious issue in the today's scenario but with an onset of mobile terminals becoming predominant, in addition to the computer security mobile security is equally noteworthy[1][2]. Mobile Phones are replaced by smart phones continually with most of the smart phones having android applications running on them. Smart Phones are used by the users for two types of functions. Firstly, the android applications on the smart phones are incorporated for personal use i.e. to obtain pictures, contacts, emails and other agendas. Secondly, the very same device is also exploited to retrieve the IT infrastructure of an organisation. The latter is implemented for policies like Bring Your Own Device. In the same scenario, there is an increase in the number of malwares attacking these devices [3].

Researchers have established two methods of malware detection. The first type is the static analysis [4] where we study malicious applications but without actually executing them. Some of the techniques used in static analysis include de-compilation, pattern matching, and decryption and so on. The unknown applications can have distinct signatures by means of obfuscation [5] and encryption therefore these methods are generally not able to identify unknown malware. Therefore, the second type of method, dynamic analysis, is proposed [6]. These approaches can monitor application's behaviours such as network access, phone calling and message sending at run time. This type of analysis is done in a sandbox environment to prevent the malware from actually infecting actual production systems; many such kinds of sandboxes are virtual systems that can be easily made to roll back to a clean state after the analysis is complete.

## II. MOTIVATION

Due to the increasing openness and popularity, android phones have been an attraction to most of the malicious applications and an attacker can easily embed its own code into the code of a benign application. Therefore, malwares attacking the android application are growing at an alarming rate and under these circumstances, security of the devices and the assets these devices allow access to, be at stake. In addition, android itself has some distinct characteristics and limitations due to its mobile nature. Therefore, detection of malware with conventional methods becomes cumbersome which poses the need to develop a novel and efficient approach for detecting malicious applications. The efficacious way of ensuring security on those devices is still a matter of research and investigation and there is a lot of room for improvement.

III. RELATED WORKS

There has been a lot of research in the field of Malware Detection. In [7], a novel dynamic analysis method named Component Traversal is proposed that can automatically execute the code routines of each given Android application (app) as completely as possible. Based on the extracted Linux kernel system calls, they further construct the weighted directed graphs and then apply a deep learning framework resting on the graph based features for newly unknown Android malware detection. However, Android apps are executed in emulator and from this data, system calls are extracted. In this scenario, some malwares are able to detect whether they run on real device or emulator and accordingly change they functionality. As a result of which, some malwares cannot be detected from this method.

In [8], a feature vector is extracted from the Android Manifest file, which combines the permission information and the component information of the Android application. Combined with the Naive Bayes classification algorithm, this approach proposes a malicious application detection method based on Android Manifest file information. This approach is a static method of malware detection which means that applications are not executed or analyzed at run time for behavioral analysis. So it cannot detect any new malware which are capable of repackaging and obfuscation to bypass their inner mechanisms.

Another approach proposed in [9] introduces a system to detect malicious Android applications through statically analyzing applications' behaviors. ANASTASIA provides a complete coverage of security behaviors. It utilizes a large number of statically extracted features from various security behavioral characteristics of an application. This approach is a static method of malware detection so it cannot protect the device from Zero Day attacks and malwares capable of modifying themselves.

Lastly [10] uses an automated feature-based static analysis method to detect malicious mobile applications on Android devices. This method uses metadata of applications and Naïve Bayes algorithm for malware detection. This approach is a static method of malware detection so it cannot protect the device from malwares that can transform themselves based on the ability to translate, edit and rewriting their own code.

IV. PROPOSED SCHEME

The approach is divided into three steps. The first step is the system call log generation. The second step is the application of the filtering method i.e. *Chi Square*. After these two steps, the dataset is created. We have managed to increase the dataset and gather the system call log information of 66 applications with each application comprising of 19 features. This combination yields the maximum accuracy. The third step of the project is the implementation of the dataset on a machine-learning algorithm. Here, we used three popular machine learning approaches i.e. Naive Bayes Algorithm, Random Forest algorithm and Stochastic Descent Gradient Algorithm. The results were obtained on the Waikato Environment for Knowledge Analysis. In addition, the Naive Bayes Algorithm and Random Forest algorithm was also implemented in C Sharp and Python respectively.

*A. System Call Log Generation*

First we install a sandbox i.e. an emulator named Genymotion (virtual device) for running each application, to prevent our own devices from getting harmed by the malicious application. Each application is executed to observe its behaviour for approximately five minutes. This results in a system call log generated using the 'strace' command. All the above is executed using python script, and figure 2 illustrates the flow of the events for the same.

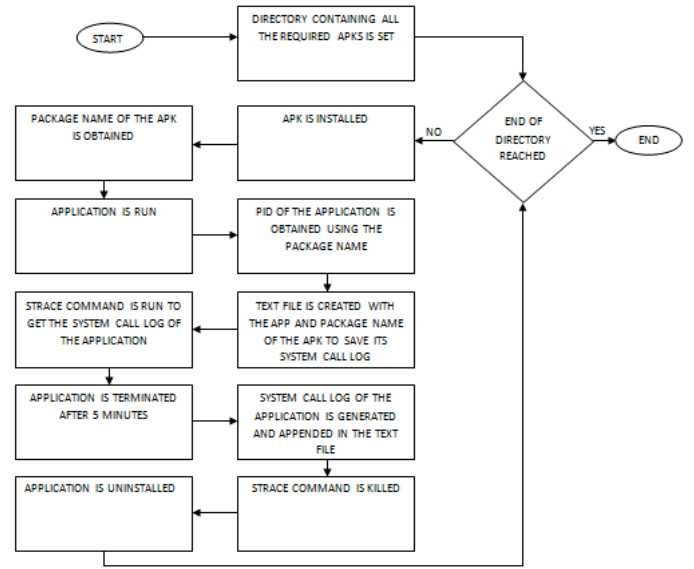

Figure 2: Flow of events for Log Generation

*B. Application if Chi-Square filtering algorithm*

Once we have the system call log information for each application, we start creating the dataset. A file consisting of system call information is created such that, corresponding to each application, the presence/absence of the system call is indicated by 1/0. Using the algorithm mentioned in Figure 3, we automate the above process. The complexity for this algorithm us $O(n)$, where n is the number of characters.

```
Algorithm 1

open excel_systemlog.txt
Declare v
Input all dataset_SystemLogs to v
n=size of v
close file
open systemlog.txt
declare set
input all app_SystemLogs to set
declare result
for i=0 to n
   begin
     if v[i] present in set
       begin
         erase v[i] from set
         result[i]=1
       end
         else result[i]=0
   end
output result
if set not empty
output set
```

Figure 3:Algorithm 1: Generation of system call binary output

After the dataset is created, to improve the accuracy for correctly identified instances, we filter the extracted features i.e. system call logs using the *chi square* algorithm. The Figure 4 clearly illustrates the algorithm we used to implement the *chi square* filtering mechanism.

```
Algorithm 2

Declare file fp
Declare ch,x,a,b,ans
While ch not equal to EOF
  begin
    if ch = 1 or ch=0
      begin
        increment count
        if ch is 0 increment y
        else if ch is 1 increment x
        if count = 11
          set b=x,d=y,x=0,y=0
        else if count=27
          set a=x,c=y
      end
  close fp
  apply chi square formula
  output ans
```

Figure 4:Algorithm 2: Chi Square filtering algorithm

As a result of which we chose the top 18 values(features), and the malware detection count added as a feature. This forms the entire dataset consisting of 66 applications and 19 features for each application. The dataset was increased to achieve this number.

### C. Machine Learning Algorithms

After the dataset is created, it is provided as an input to three algorithms. The first algorithm is Naive Bayes Algorithm [11] which is implemented using C sharp language. In the same, first the training set (a part of the data set just created) is stored in Microsoft Access (used for database connectivity and storage). The rest of the part is stored in another table in order to gain the accuracy of unknown samples. After this, for each feature, mean and variance is computed. These values help us construct a classifier table which is used to classify the unknown apps. Each row of the second table acts as an unknown sample and results obtained provide us with the accuracy. Figure 5 illustrates the flow of events of Naive Bayes Classifier.

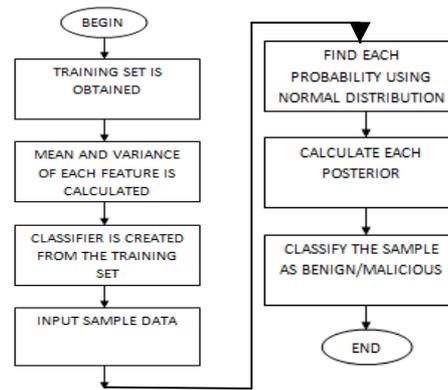

Figure 5: Naive Bayes Classifier

The same dataset is then input to a second classification algorithm i.e. Random Forest. In Random Forest, the dataset is first created and is divided into random subsets. Then a decision tree is created for each subset. On the basis of their results, the trees are grouped. To classify the unknown sample as benign or malicious, we use the group having maximum number of trees.

Figure 6 shows the flow of this algorithm.

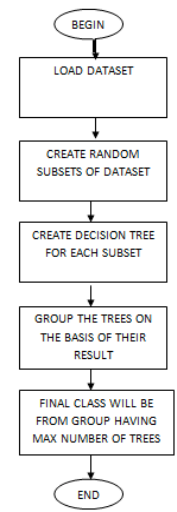

Figure 6: Random Forest Classifier

The dataset was then input on the SGD algorithm as well to obtain the results. After the implementation, we have managed to achieve an accuracy of 93.75% on Naive Bayes Classifier and 93.84% on Random Forest Classifier [12] and 95.5% on SGD. These results were first obtained on the Weka explorer and then verfied by the implementation using C Sharp and Python Language. The results are illustrated in Figure 7

### V. RESULTS AND IMPLEMENTATION

After the implementation of the dataset on these algorithms, we calculate the true positive value, false positive value, accuracy, precision, recall, and F measure. Table 2 shows the methods to calculate all the values.

**Table 2: Important Parameters and their Calculations Methods**

| Parameter | Calculation Method |
|---|---|
| True Positive (TP) | A true positive result is the one that detects the condition(presence of malware) when the condition is present. |
| True Negative (TN) | A true negative result is the one that does not detect the condition(presence of malware) when the condition is absent. |
| False Positive (FP) | A false positive result is the one that does detect the condition(presence of malware) when the condition is absent. |
| False Negative (FN) | A false negative result is the one that does not detect the condition(presence of malware) when the condition is present. |
| Precision (p) | TP/(TP+FP) |
| Recall (r) | TP/(TP+FN) |
| F Measure | (2*p*r)/(p+r) |
| Positive Predictive Value | TP/(TP+FP) |
| Negative Predictive Value | TN/(TN+FN) |
| Accuracy | TP+TN/(total number of records to be tested) |

Table 3, 4 and 5 illustrate the calculated values on the three algorithms.

**Table 3: Important Parameters and their Calculations (Naive Bayes)**

| Parameter | Calculated Value |
|---|---|
| True Positive Rate (TP) | .955 |
| False Positive Rate (FP) | .080 |
| Precision (p) | 0.958 |
| Recall (r) | .955 |
| F Measure | 0.954 |

**Table 4: Important Parameters and their Calculations (Random Forest)**

| Parameter | Calculated Value |
|---|---|
| True Positive Rate (TP) | .985 |
| False Positive Rate (FP) | .010 |
| Precision (p) | 0.985 |
| Recall (r) | .985 |
| F Measure | 0.985 |

**Table 5: Important Parameters and their Calculations (Random Forest)**

| Parameter | Calculated Value |
|---|---|
| True Positive Rate (TP) | .955 |
| False Positive Rate (FP) | .056 |
| Precision (p) | 0.955 |
| Recall (r) | .955 |
| F Measure | 0.954 |

We applied the dataset on three different algorithms, and calculate the accuracy and compared the values. Figure 7 shows the comparison between these algorithms. From the figure, we can clearly justify the quality of the dataset since all the algorithms are able to identify a considerable number of instances.

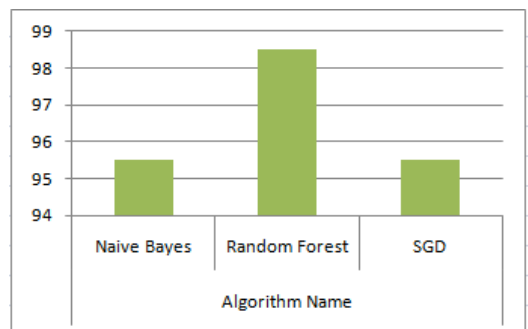

**Figure 7: Implementation Results**

On the basis of the results obtained above, i.e. the high accuracies in the three algorithms, we can say that correctness of the dataset formed is justified.

## VI. CONCLUSION

This paper defines an approach that creates a dataset using system call log information. The dataset is then implemented on three algorithms namely, Naive Bayes algorithm, Random Forest Algorithm and Stochastic Gradient Descent algorithm. The results are analysed and accuracy is calculated. The main features of our algorithm include: Firstly, usage of system call logs i.e. working at the kernel level to find the malicious behaviour of the applications. Secondly, the entire process is automated consuming a minimum amount of time. Thirdly, the quality of the dataset is improved by using a filtering algorithm called *Chi Square*. Lastly, the correctness and quality of the dataset is justified with the high accuracy results we obtained. The future scope of the work includes identification of those applications, which are capable of identifying a sandbox type environment and behaving accordingly. Under such circumstances, sandboxes like Genymotion cannot be used.